\documentclass[aps,prl,twocolumn,groupedaddress,showpacs,superscriptaddress]{revtex4-1}
\usepackage{aas_macros}

\usepackage{textcomp}
\usepackage{gensymb}
\usepackage{graphicx}
\usepackage{dcolumn}
\usepackage{bm}
\usepackage[colorlinks,citecolor=blue]{hyperref}

\usepackage{color}

\begin{document}

\preprint{APS/123-QED}

\title{Self-lensing flares from BH binaries II:\\ observing black hole shadows via light curve tomography}

\author{Jordy Davelaar}
 \email{jrd2210@columbia.edu}
\affiliation{Department of Astronomy, Columbia University, 550 W 120th St, New York, NY 10027, USA}
\affiliation{Columbia Astrophysics Laboratory, Columbia University, 550 W 120th St, New York, NY 10027, USA}%
\affiliation{
 Center for Computational Astrophysics, Flatiron Institute, 162 Fifth Avenue, New York, NY 10010, USA}%

\author{Zolt\'an Haiman}
\affiliation{Department of Astronomy, Columbia University, 550 W 120th St, New York, NY 10027, USA}

\date{\today}
\begin{abstract}
Supermassive black hole (BH) binaries are thought to produce self-lensing flares (SLF) when the two BHs are aligned with the line-of-sight. If the binary orbit is observed nearly edge-on, we find a distinct feature in the light curve imprinted by the relativistic shadow around the background ("source") BH. We study this feature by ray-tracing in a binary model and predict that 1\% of the current binary candidates could show this feature. Our BH tomography method proposed here could make it possible to extract BH shadows that are spatially unresolvable by high-resolution VLBI.
\end{abstract}

\keywords{Black hole binaries - gravity - radiative transfer}
\maketitle

\section{Background}
Supermassive black hole binaries (SMBHBs) are thought to reside in the cores of many galaxies, as a result of galaxy mergers \citep{begelman1980}. Their immediate surroundings consist of a circumbinary disk that transfers material to two minidisks, each orbiting one BH component. If the viewing angle with respect to the orbital plane is close to edge-on, a minidisk can get lensed by the foreground BH resulting in a self-lensing flare (SLF) \citep{dorazio2018,ingram2021}, especially in the case of ultra-compact binaries close to merger, in the regime where their gravitational wave (GW) emission is detectable by LISA~\cite{haiman2017}.

Observational evidence of SLFs is sparse, a first candidate was identified by \cite{hu2020} in the optical {\it Kepler} data, KIC-11606854 (dubbed "Spikey"). A second candidate, identified in X-rays, was discussed by ~\cite{ingram2021}. More generally, quasi-periodic modulations in the light curve of AGN indicating the presence of a SMBHB have recently been found in large optical time-domain surveys \citep{graham2015,charisi2016,liu2019,liu+2020,chen+2020}.

The first attempts of modeling SLFs were made by using a point source approximation for the lens and the source, with the amplification factor derived from microlensing \citep{haiman2017,dorazio2018,hu2020}. These models are computationally cheap, but they lack general relativistic (GR) effects, and do not take finite source sizes into account. Ref.~\citep{dorazio2018} also studied the impact of finite source size, but their emission morphology, lacks the strong bending of light close to the BH since GR is not taken into account. Refs.~\cite{ingram2021} and \cite{pihajoki2018} used general-relativistic ray tracing (GRRT) to study SLFs, but this was either limited to a single mass ratio q=0.01 \cite{pihajoki2018} or aimed at a single source and only considered circular equal-mass binaries \cite{ingram2021}. \cite{pihajoki2018} used a superimposed binary metric, while \cite{ingram2021} used an image of a BH as a faraway image lensed by a single Kerr BH. See also for various applications of lensing in compact binaries \citep{jaroszynski1992,beky2013,bohn2015,kelly2017,dascoli2018,schnittman2018}.

In single BH emission models, there is typically a flux depression present in the apparent image. In the optically thin case, this "hole" in the image coincides with the BH shadow (BHS) \cite{luminet1979,falcke2000b,bronzwaer2021}, while in the optically thick case, the position of the lensed event horizon or the inner-shadow \cite{gralla2019,chael2021}. The shadow is a flux depression in the apparent surface brightness, caused by photons being trapped by the event horizon. The first direct observation of such a flux depression or BHS was made by the Event Horizon Telescope (EHT) \cite{eventhorizontelescopecollaboration2019}. They used very-long baseline interferometry to construct an image of the immediate surrounding of the SMBH in the nucleus of M87. A limitation of direct imaging is that extremely high spatial resolution of $20 ~\mu$-arcseconds was needed to resolve M87*. This limits the potential number of BHs measurable by the current EHT array, although this number is expected to grow with future EHT upgrades \citep{johnson2019,pesce2021}.

In \cite[][hereafter Paper~I]{davelaar2021} we present a comprehensive suite of SLF models, obtained via GRRT calculations with our modified version of the GRRT code {\tt RAPTOR} \cite{bronzwaer2018,bronzwaer2020}. The code includes a superposed binary metric where the BHs are on Keplerian orbits. The emission is generated by two Novikov-Thorne like \citep{novikov1973} minidisks. Each minidisk extends from the horizon or the innermost stable circular orbit (ISCO) to the tidal truncation radius. For full details on the code and model, see Paper~I.

In this paper, we investigate substructure found in the SLFs for a subset of models in Paper~I. These "dips" occur when the BHS passes behind the lens.
We focus on high-energy emission, which is concentrated in a small region around each BH component, and shares the BH orbital motions, even for the compact binaries of interest~\cite{haiman2017}.
We generate light curves at 2.5 keV for model parameters affecting
the shape of the dip: the inner radii of the minidisks $R_{\rm inner}$ \footnote{Changing the inner radius also changes the velocity profile, Keplerian if at the ISCO or including a radial component if at the event horizon.}, the binary's separation $a_{\rm orb}$, BH spins $a_{\rm BH}$, disk opacity $\tau$, and
binary inclination $i_{\rm orb}$. The model parameters are listed in Table~\ref{tab:params}.

\begin{table}
\begin{tabular}{l|lllll}
 & $R_{\rm inner}$ &  $a_{\rm orb}$ & $i_{\rm orb}$  & $a_{\rm BH}$ &  $\tau$    \\
  \hline
M0 & $r_{\rm h}$ & 100 &  90 & 0 & thick    \\
\hline
M1 & \bm{$r_{\rm ISCO}$} & 100 &  90 & 0 & thick    \\
\hline
M2 & $r_{\rm h}$ & 100 &  90 & 0 & \bf{thin}    \\
\hline
M3a & $r_{\rm h}$ & 100 &  90 & \bf{0.5} & thick    \\
M3b & $r_{\rm h}$ & 100 &  90 & \bf{0.95} & thick    \\
\hline
M4a & $r_{\rm h}$ & \bf{200} &  90 & 0 & thick    \\
M4b & $r_{\rm h}$ & \bf{300} &  90 & 0 & thick    \\
M4c & $r_{\rm h}$ & \bf{400} &  90 & 0 & thick    \\
M4d & $r_{\rm h}$ & \bf{500} &  90 & 0 & thick    \\
M4e & $r_{\rm h}$ & \bf{1000} &  90 & 0 & thick    \\
\hline
M5a & $r_{\rm h}$ & 100 &  \bf{89} & 0 & thick    \\
M5b & $r_{\rm h}$ & 100 &  \bf{88} & 0 & thick    \\
M5c & $r_{\rm h}$ & 100 &  \bf{87} & 0 & thick    \\
M5d & $r_{\rm h}$ & 100 &  \bf{86} & 0 & thick    \\
M5e & $r_{\rm h}$ & 100 &  \bf{85} & 0 & thick    \\
M5f & $r_{\rm h}$ & 100 &  \bf{80} & 0 & thick    \\
\hline
M6a & $r_{\rm h}$ & \bf{200} &  \bf{89} & 0 & thick    \\
M6b & $r_{\rm h}$ & \bf{200} &  \bf{88} & 0 & thick    \\
M6c & $r_{\rm h}$ & \bf{200} &  \bf{87} & 0 & thick    \\
M6d & $r_{\rm h}$ & \bf{200} &  \bf{86} & 0 & thick    \\
M6e & $r_{\rm h}$ & \bf{200} &  \bf{85} & 0 & thick    \\
M6f & $r_{\rm h}$ & \bf{200} &  \bf{80} & 0 & thick    \\
\end{tabular}
\caption{Model parameters. M0 is the fiducial model. In each variant, a single parameter is varied, shown in bold.}
\label{tab:params}
\end{table}

\section{Fiducial self-lensing flare}
\begin{figure}[t]
  \centering
  \includegraphics[width=0.5\textwidth]{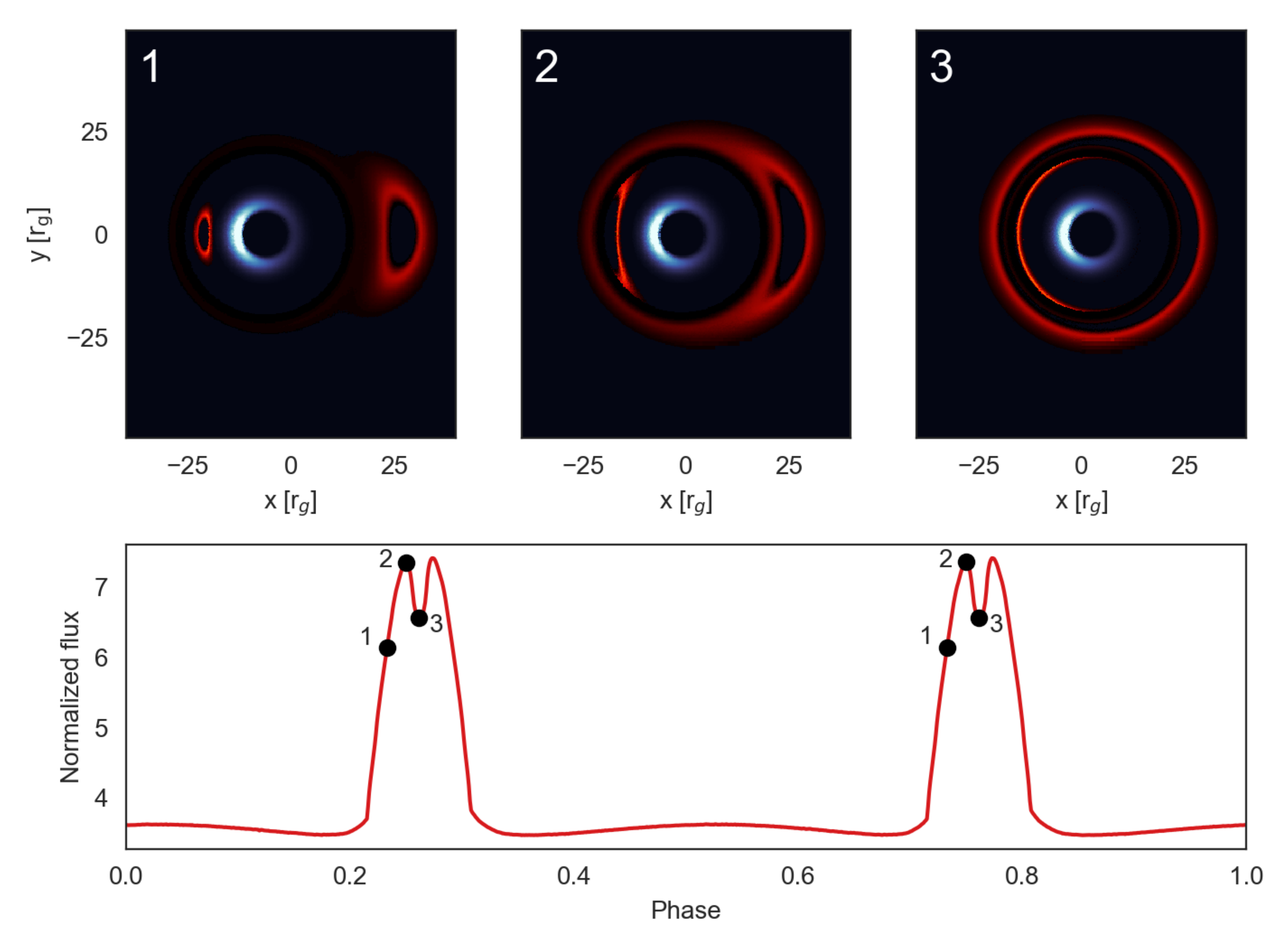}
  \caption{Fiducial model. The top row shows three snapshots during the self-lensing flare. The bottom row shows the light curve at 2.5 keV. Numbers shown on the light curve correspond with numbers in the top panels.}
  \label{fig:fid}
\end{figure}

In Figure~\ref{fig:fid} we show our fiducial model M0. The bottom panel shows the light curve at 2.5 keV that contains two SLFs at a quarter and three-quarters of the orbit. The SLFs have a distinct dip at their peak, which occurs when the BHS passes the lens' focal point, resulting in a drop in the net amplified flux. The top panels of Figure~\ref{fig:fid} show synthetic images of model M0 at the start (1), peak (2), and dip (3) of the SLF. A secondary image is visible on the left.

\section{Parameter dependencies}
\begin{figure}[t]
    \includegraphics[width=0.5\textwidth]{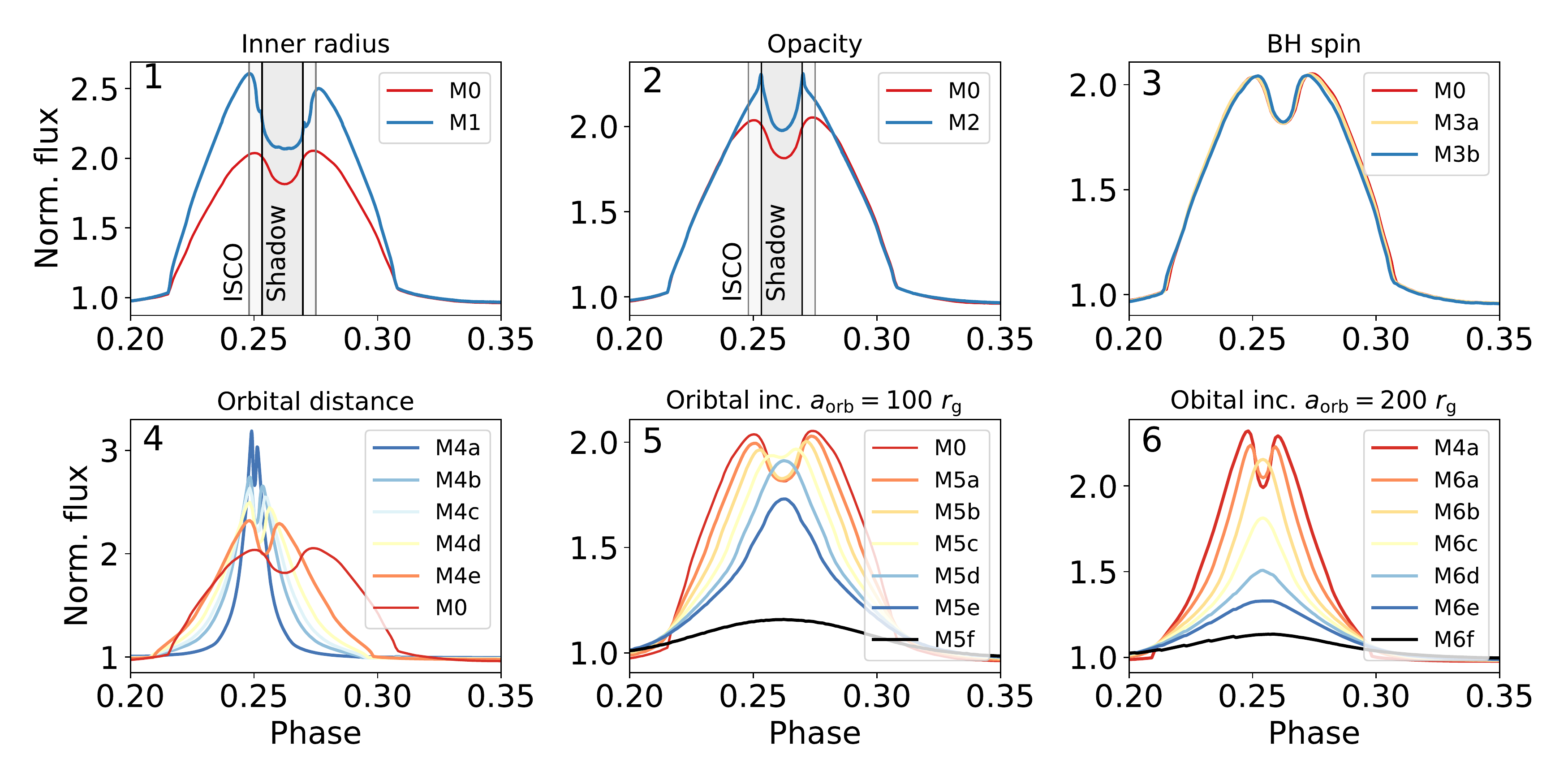}
  \caption{Various light curves are shown at 2.5 keV for different model parameters.
  {\it Panel~1}: dependence on the {\it inner radius}, which is at the horizon in model M0 and at ISCO in model M1. Vertical black lines (also in Panel~2) mark the time interval corresponding to the BHS size and gray lines the ISCO.
  {\it Panel~2}: dependence on the minidisk {\it optical depth}, which is assumed to be thick in model M0, and thin in model M2.
  {\it Panel~3}: dependence on the BH {\it spin parameter}, which is assumed to be $a=0, 0.5$ and 0.95 (the same for both BHs) in models M0, M3a, and M3b, respectively.
  {\it Panel~4}: dependence on {\it binary separation}, which is $100$, $200$, $300$, $400$, $500$ and $1000~R_{\rm g}$ in models M0 and M4a-e, respectively.
  {\it Panel~5}: dependence on {\it inclination} of the binary orbital plane, which is $90^{\degree}$(edge-on),  $89^{\degree}$, $88^{\degree}$, $87^{\degree}$, $86^{\degree}$, $85^{\degree}$ and $80^{\degree}$ in models M0 and M5a-f, respectively.
  {\it Panel~6}: same as Panel~5, but for a binary separation of $200~R_{\rm g}$.}
  \label{fig:multi}
\end{figure}

In model M1, we increase the inner radius, which widens the spacing between the sub-peaks, see Panel~1 of Figure~\ref{fig:multi}. Additionally, since the truncation radius is larger than the photon ring, a second set of sub-flares is visible. The sub-flares have a phase spacing of $\Delta \phi \approx 0.0165$.  Multiplying this value by the total circumference of the orbit ($628 ~R_{\rm g}$, where $R_{\rm g}\equiv GM/c^2$ is the gravitational radius), we find that this phase difference corresponds to a size of $d \approx 10.4 ~R_{\rm{ g}}$, which is the size of the projected shadow. For guidance, the vertical lines mark the phase duration of the ISCO and photon rings centered on the SLF.

In the optically thin case, model M2, presented in Panel 2 of Figure~\ref{fig:multi}, the main sub-peaks trace the photon ring, since this is the dominant feature in the image. The spacing in phase between the sub-peaks
is equal to that for the sub-peaks produced by the photon ring in model M1. We again mark the phase durations of the ISCO and photon rings centered on the middle of the SLF. The dip inside the flare coincides in phase with the shadow's passage behind the lens.

In models M3a-b, we vary the spin of the BHs (assumed to be the same for both components for simplicity) to $a_{\rm BH}=0.5$ and $0.95$, respectively, compared to non-spinning fiducial case. The light curves of these models are shown in Panel 3 of Fig. \ref{fig:multi}. The addition of spin results in a slightly smaller and more asymmetric shadow. This is also visible in the light curves; the spacing between the peaks shrinks for increasing spin. However, as the figure shows, the effect of spins is overall very modest.

In models M4a-f we study the dependence on binary separation by changing $a_{\rm orb}$ to $200$, $300$, $400$, $500$ and $1000~R_{\rm g}$, compared to $100~R_{\rm g}$ in the fiducial model. Increasing the separation results in the light curves asymptote to the point source approximation, since the angular size of the source on the sky decreases. Additionally, the source spends a smaller fraction of the orbit within the Einstein radius.
In M5a-f we alter the inclination to $89^{\degree}$, $88^{\degree}$, $87^{\degree}$, $86^{\degree}$, $85^{\degree}$ and $80^{\degree}$ respectively, compared to the fiducial  $90^{\degree}$. These light curves are shown in Panel~5 of Fig.~\ref{fig:multi}.  The dip is clearly present in models M5a-c, which puts a limit on the range of inclinations for which it can be observed, $i_{\rm orb} = 90^{\degree} \pm 3^{\degree}$. In model M6a-f we increase the binary separation to $200 ~R_{\rm{g}}$ and cover the same span of inclinations. The light curves of these models can be seen in Panel~6 of Fig.~\ref{fig:multi}. As the binary separation increases, the inclination window shrinks to $i_{\rm orb} = 90^{\degree} \pm 2^{\degree}$.

\section{Analytic expectations}
In the case of a perfectly edge-on circular binary, we can derive the expected phase spacing between the two peaks in the optically thin case by taking the ratio between the diameter of the shadow and the circumference of the orbit, %
\begin{equation}
    \Delta \phi = \frac{d_{\rm{shadow}}}{2 \pi a_{\rm orb}},
    \label{eqn:1}
\end{equation}
where $a_{\rm orb}$ is the orbital radius, and {$d_{\rm shadow} = 2 \sqrt{27} G M_{\rm source}/c^2$} is diameter of the BHS of the source \citep{hilbert1917,bardeen1973,luminet1979,falcke2000b}, where $M_{\rm source}$ is related to the full binary mass via $M_{\rm bin} =  q M_{\rm source}/(1+q)$. This assumes the source is the secondary BH, which in general is expected to out-accrete and out-shine the primary BH~\citep{farris2014,duffell2020}.

The orbital radius is given by $a_{\rm orb} = \sqrt[3]{\frac{4\pi^2 T^2}{G M_{\rm bin}}}$, where $T$ is the orbital period and $M_{\rm bin}=M_1+M_2$ is the total binary mass. Combined with Eq.~\ref{eqn:1} and using geometrized units ($G=c=1$), yields
\begin{equation}
    \Delta \phi = \frac{2 \sqrt{27}}{(4 \pi^2)^{1/6}} \frac{q M_{\rm bin}^{2/3}}{(1+q) T^{2/3}} \approx 5.63 \frac{q M_{\rm bin}^{2/3}}{(1+q) (T)^{2/3}}
\end{equation}
For our fiducial model, $M_{\rm bin} =2\times10^7~{\rm M_\odot}$, $q=1$ and $T=4443 ~R_{\rm g}/c$, which gives $\Delta\phi = 0.0165$, identical to what we found from our light curves. We note that this relation is less trivial for an eccentric binary, since the velocity depends on the nodal angle at which the BHs align and produce the SLF.

\begin{figure}[htp]
  \centering
  \includegraphics[width=0.5\textwidth]{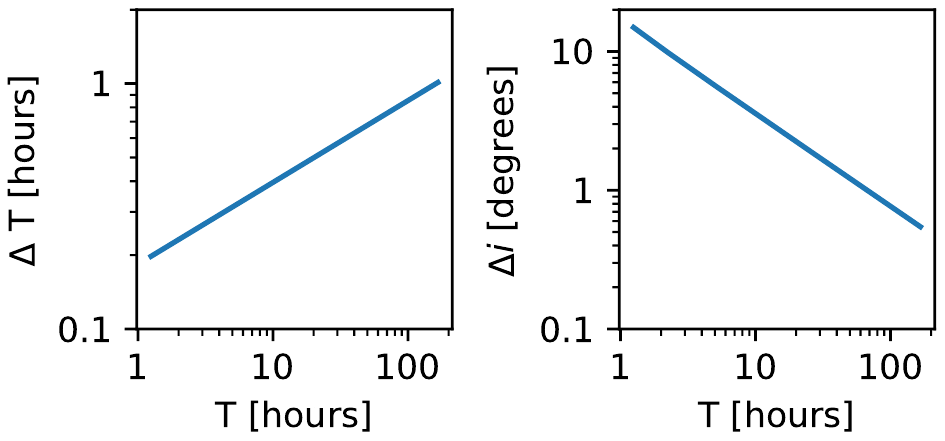}
 \caption{The {\it left panel} shows the spacing between the two sub-peaks imprinted on the light curve by the BH shadow as a function of the orbital period for a circular equal-mass $M_{\rm bin}=2 \times 10^7 ~M_\odot$ binary.
 The {\it right panel} shows the maximum tilt from an edge-on view for which a double-peaked SLF peak structure appears.}
    \label{fig:analytical}
\end{figure}

Next, we derive an expression for the range of inclination angles $\Delta i$. The dip is visible when the focal point of the lens moves over the shadow. This requires that the inclination does not exceed the angular size of the shadow on the sky, or $\Delta i=\sin^{-1}(d_{\rm shadow}/(2 a_{\rm orb}))$.  Using Eq.~\ref{eqn:1} we then find
\begin{equation}
    \Delta i = \sin^{-1}( \pi \Delta \phi).
\end{equation}
Inserting the fiducial model values, we find $\Delta i = 3^{\degree}$, confirming our numerical result from Panel~5 of Fig.~\ref{fig:multi}. In Figure~\ref{fig:analytical} we show the relations for the time interval $\Delta T = \Delta \phi T$ and the inclination window $\Delta i$ for a binary with mass $M_{\rm bin}=2\times10^7 M_\odot$ where we vary the period. As $T$ increases, which is equivalent to increasing the binary separation, the time interval increases, and the inclination window decreases. The spacing of the dip in our fiducial model is $\Delta \phi  = 0.0165$. Converting this to physical time for a binary with mass $M_{\rm bin} =2\times 10^7 M_\odot$, we find a spacing of 30 minutes. This puts a limit on the time cadence required by an observation (although repeated observations could be phase-folded and reveal the dip in a more coarsely sampled flare). As a function of mass, the orbital period $T$ and $\Delta T$ both scale linearly with mass, making more massive binaries easier to resolve in time.

\section{Observational limitations}
Ideally, individual SLFs should outshine the accretion-induced variability and instrumental noise~\footnote{In principle, the periodically recurring SLFs could be discovered via a blind phase folding of the light curve, even if they are individually below the noise.}.
Typical amplitudes of stochastic AGN variability on timescales of hours to days are 10-40 \% in the X-rays \citep{lawrence1993,maughan2019}, and 5--10\% in the optical on timescales of months \citep{peterson2001}. From the inclination-dependent models M25a-f, we find that the flare amplitude exceeds 20\% of the unlensed flux when the inclination is larger than $80^{\degree}$. This inclination window for producing 20\% flares widens/narrows for more compact/wider binaries (see Fig.~\ref{fig:analytical}).
The depth of the dip is itself $\approx 20$\% in the fiducial model, and reduces as the separation increases. The duration of the dip is on the order of hours to days depending on BH mass. The flux variability in AGN typically shows a stochastic red-noise power spectrum \citep{done1992,mushotzky1993,vaughan2003,panagiotou2020}. This is beneficial since the amplitude of the variability on the short timescales of the SLFs is lower. Additionally, phase-folding can be used to average out stochastic variability.

Large time-domain surveys discovered dozens of candidate SMBHBs based on periodicity in optical light curves~\citep{graham2015,charisi2016}, although with sparsely sampled light curves. The estimated BH mass distribution of these candidates is skewed towards the higher BH masses of $10^{9-10} M_\odot$. This widens the dip's duration from an  $\sim$hour in our fiducial model to a few days (scaling only the BH mass). By increasing the period of the orbit, this interval would further widen (see Fig.~\ref{fig:analytical}) and exceed the five-day cadence for some candidates in \citep{charisi2016}.

There are some caveats to the detectability of SLFs and their "dips".  If the image of the source is highly asymmetric, e.g., strongly Doppler-deboosted on one side of the image, the secondary peak we observe in the light curve will be less prominent~(Paper~I). Using a physical model for the accretion flow, e.g. from hydrodynamic or GR magnetohydrodynamic (GRMHD) simulations, should help clarify the visibility of these features. A recent work~\cite{dascoli2018} ray-traced the emission from a binary in GRMHD simulations, and examined the time-averaged bolometric flux as a function of the azimuthal position of an observer at different latitudes. This is an estimate of the phase-folded light curve, and a hint of a dip is visible for nearly edge-on viewing angles (their Fig.~11). The deleterious effects of asymmetric Doppler boosting could also be mitigated if the plane of one (or both) minidisks is misaligned with the binary's orbital plane. Such misalignment, or related lateral tearing of the minidisks, is possible, if the minidisk orbital angular momentum and BH spin vectors are aligned \citep{bardeen1975,nixon2011,liska2019}, although this configuration is expected to be less common~\citep{Gerosa+2015, Nealon+2021}.

Another possible limitation is that the edge-on circumbinary disk can block the view to the event horizon. However, the binary orbital plane can be misaligned with respect to the circumbinary disk \cite{nixon2011,nixon2011b,hayasaki2013}. Furthermore, for thin AGN disks with accretion rates near but below the Eddington limit, with a seperation of $\approx 100$ $R_{\rm g}$, the disk aspect ratio is $\lesssim 0.03$~\cite{haiman+2009}, so that the circumbinary disk would not obscure the flare from a coplanar binary.

Finally, compact binaries exhibiting SLFs may be rare because they are short-lived. For instance the GW-driven inspiral time in our fiducial model is a mere 1.5 years (\citep{peters1964}, see Eq.~28 in \citep{haiman+2009}). This is mitigated for larger BH masses since the inspiral time scales linearly with BH mass. For example, a $10^9 ~M_\odot$ BH would increase the inspiral time by a factor hundred. Additionally, wider binaries are substantially more long-lived since the inspiral time scales with $a_{\rm orb}^4$. Wider orbits, which also have wider dips, should ensure that a substantial population is present that might have the right condition for future observations to detect SLFs with BH shadows.

\section{Search strategies}
An estimation of the probability of observing SLFs was performed by \citep{dorazio2018}, who estimated that 10\% of the currently known SMBHB candidates potentially produce SLFs. These candidates typically have orbital periods of a year and masses of $\sim 10^9 M_\odot$. The visibility of the dip is strongly limited by the orbital inclination, and decreases the probability to observe it. From our formula we find, for the above high-mass candidates, a window of $\Delta i \approx 1^{\degree}$, corresponding to a 1\% chance assuming isotropic inclinations. Since 150 candidates are known, one of these could be in the right inclination regime.

Since the probability of detection is low, we propose two strategies to find SLFs with BLS features. First, SLFs could be observable by current and future optical and near-infrared ground-based telescopes, designed for time-domain surveys. One option would be to perform follow-up observations of the already known SMBHB candidates with a $\sim$daily cadence. Additionally, the Rubin Observatory's LSST is expected to identify 20-100 million quasars \citep{VRO2009,xinhaiman2021} down to a BH mass of a few $\times 10^{5} ~{\rm M_\odot}$. The sheer number of potential sources increases the probability of finding AGNs with the right conditions to measure BHSs, Ref. \citep{kelley2021} estimates that hunderds of potential SLFs could be found. The light curves of these quasars will also be sampled at a cadence of a few days, which should be suitable to identify many periodic sources and SLFs.

A second strategy is to perform electromagnetic followups of on-going SMBHB mergers discovered with LISA~\citep{LISA}. LISA is expected to detect SMBHBs at their late inspiral phase starting at periods of a few hours. At this moment, the binary is already compact (separation $\sim100~R_{\rm g}$~\cite{haiman2017}), which is favorable for observing the BHS since the inclination window increases with decreasing binary period (see Fig. \ref{fig:analytical}). LISA binaries can be identified and localized on the sky to several square degrees 1-2 days prior to their merger~\cite{mangiagli+2020}, allowing an EM search with wide field-of-view telescopes covering dozens of their orbits. Accurate knowledge of the orbital phase from the GW data will facilitate a concurrent search for SLFs.

\section{Summary}
We presented numerical models for SLFs in the light curves of SMBHBs, based on GRRT simulations of thin disk emission in an approximate binary metric. If such a SMBHB system is observed close to edge-on, we find a double-peaked substructure in the SLF, where the spacing of the sub-peaks is set by the angular size of the BHS. For our fiducial model the probability of detecting this dip is $\approx 3$\%, assuming isotropic inclinations. The BH tomography method proposed here would open a new window to discovering BHSs and characterizing their size. The features in the SLFs  also yield independent constraints on the BH masses, properties of their emission morphology and kinematics, and the binary's orbital parameters. Since this method does not require spatially resolving the source, it allows probing BHs which are out of reach for the EHT.

\begin{acknowledgments}
{\it Acknowledgments -- } The Authors thank Christiaan Brinkerink, Anastasia Gvozdenko, Jeremy Schnittman, and Daniel D'Orazio for valuable comments and discussions during this project. JD is supported by a Joint Columbia/Flatiron Postdoctoral Fellowship. Research at the Flatiron Institute is supported by the Simons Foundation. We acknowledge support by NASA grant NNX17AL82G and NSF grants 1715661 and AST-2006176. Computations were performed on the {\tt Popeye} computing cluster at SDSC maintained by Flatiron Institute's SCC. This research has made use of NASA's Astrophysics Data System.
{\it Software:} {\tt python} \citep{oliphant2007,millman2011}, {\tt scipy} \citep{jones2001}, {\tt numpy} \citep{vanderwalt2011}, {\tt matplotlib} \citep{hunter2007}, {\tt RAPTOR} \citep{bronzwaer2018,bronzwaer2020}.
\end{acknowledgments}

\end{document}